\documentclass{PoS}

\title{Wilson fermions in the epsilon regime}

\ShortTitle{Wilson fermions in the epsilon regime}

\author{Oliver B\"ar\\
Institut f\"ur Physik, Humboldt Universit\"at, 
Newtonstrasse 15, 12489 Berlin, Germany\\
E-mail: \email{obaer@physik.hu-berlin.de}}

\author{\speaker{Silvia Necco}\\
        CERN, Physics Departement, 1211 Geneva 23, Switzerland\\
        E-mail: \email{Silvia.Necco@cern.ch}}

\author{Stefan Schaefer\\
Institut f\"ur Physik, Humboldt Universit\"at, 
Newtonstrasse 15, 12489 Berlin, Germany\\
E-mail: \email{sschaef@physik.hu-berlin.de}}

\abstract{
We extend the $\epsilon$-expansion of continuum chiral perturbation theory to nonzero lattice spacing in the framework of Wilson Chiral Perturbation Theory.
We distinguish various regimes by defining the relative power counting of the quark mass $m$ and the lattice spacing $a$. We observe that for $m\sim a\Lambda^2_{\rm QCD}$, the explicit breaking of chiral symmetry in Wilson fermions is still driven by the quark mass and lattice corrections are highly suppressed. 
The lattice spacing effects become more pronounced for smaller quark masses and may lead to non-trivial corrections of the continuum results at next-to-leading order. We compute these corrections for standard current and density correlation functions. A fit to lattice data shows that these corrections are small, as expected. 

\vspace{2cm}
\begin{flushright}
HU-EP-09/49\\
CERN-PH-TH-2009-187
\end{flushright}
}

\FullConference{The XXVII International Symposium on Lattice Field Theory\\
        July 26-31, 2009\\
        Peking University, Beijing, China}

\begin{document}

\section{Introduction}
Matching results of light quarks lattice QCD simulations with the predictions of the chiral effective theory is an important step towards a comprehensive picture of low energy properties of strong interactions.  
Through this matching one can eventually extract the Low-Energy Couplings (LECs) of the effective theory: many results have been presented in the past months for the leading order (LO) and next-to-leading order (NLO) couplings, both for the SU(2) and the SU(3) effective theory (for a recent review see for instance \cite{Necco:2009cq}). 
While lattice QCD simulations in the chiral regime experience constant progress, the control over the systematic uncertainties remains a relevant issue. Lattice artifacts and finite-volume effects must be thoroughly understood. Moreover, since the matching is done at a given order in the chiral effective theory, the effect of the neglected higher order contributions must be under control.

For this reason, it is very useful to extract the LECs from a large set of independent observables and from different kinematic regimes.
An interesting approach is to investigate QCD in a finite volume $V=L^3 T$ in the so-called $\epsilon$-regime \cite{Gasser:1986vb}, where the pion wavelength is larger than the size of the box, $M_\pi L< 1$. In this situation volume effects are enhanced, 
while mass-effects are suppressed with respect to the usual infinite-volume case (or $p$-regime, where $M_\pi L\gg 1$). The consequence is that, at a given order in the perturbative expansion, less unknown LECs will appear, making the $\epsilon$-regime potentially convenient for the extraction of the LO constants.

Lattice simulations in the $\epsilon$-regime are challenging, since one still needs a fairly large volume ($L\gg 1/(4\pi F)$), while the quark mass must be very small ($m\Sigma V \lesssim 1$). This fact strongly influences the choice of the lattice action to be used.
Ginsparg-Wilson (GW) fermions \cite{Ginsparg:1981bj} have many theoretical advantages, since they guarantee exact chiral symmetry at finite lattice spacing \cite{Luscher:1998pq}. 
The price to pay to maintain continuum-like chiral symmetry is the high computational cost. While a large number of quenched computations in the $\epsilon$-regime with Ginsparg-Wilson Dirac operator has been performed (see \cite{Giusti:2008fz} for a recent study, and references therein for precedent computations), dynamical simulations are still very costly. For recent calculations with $N_f=2$ dynamical chiral fermions see \cite{Fukaya:2007pn,DeGrand:2007tm,Hasenfratz:2007yj,Lang:2006ab}. H. Fukaya presented at this conference new preliminary results for $N_f=2+1$ obtained by JLQCD/TWQCD \cite{fukaya_plenary}. 

On the other hand, dynamical simulations with the Wilson Dirac operator with $O(a)$ improvement are becoming fairly inexpensive. 
Nevertheless, since chiral symmetry is explicitly broken at finite lattice spacing, the spectral gap is not bounded from below: at small quark masses the probability distribution of the gap can approach zero values, leading to possible integration instabilities and sampling inefficiencies in the algorithm. 
In \cite{DelDebbio:2005qa} the empirical stability bound $m>m_{min}$, with $m_{min}\propto a/\sqrt{V}$ has been established. A possible solution to this problem comes from a reweighting technique \cite{Hasenfratz:2008fg} (see also \cite{Luscher:2008tw}). This method has been applied in \cite{Hasenfratz:2008ce} to simulate Wilson fermions in the $\epsilon$-regime. Lattice simulations in the $\epsilon$-regime are feasible also with Wilson Twisted Mass fermions, as presented in \cite{Jansen:2007rx} and also at this conference \cite{shindler_lat09}.\\
In principle, the matching with the chiral effective theory should be performed only after a continuum extrapolation of the lattice results. While this is not an unrealistic goal for the near future, the presently available simulations in the $\epsilon$-regime are carried out at a single value of the lattice spacing.  
In \cite{Hasenfratz:2008ce} the pseudoscalar and axial correlations functions turned out to be very well described by continuum chiral effective theory at NLO. Still, it is important to have a theoretical understanding of the impact
of explicit breaking of chiral symmetry on computations in the $\epsilon$-regime. 
We address this question in \cite{Bar:2008th}:the tool that we adopt is the so-called Wilson Chiral Perturbation Theory (WChPT) \cite{Sharpe:1998xm,Rupak:2002sm}, the low-energy effective theory for lattice QCD with Wilson Dirac operator. A similar analysis has been carried out in \cite{Shindler:2009ri,shindler_lat09}.

\section{Wilson Chiral Perturbation Theory}
The chiral effective Lagrangian of WChPT is an expansion in powers of pion momenta $p^2$, the quark mass $m$ and lattice spacing $a$. Based on symmetries of the underlying Symanzik action \cite{Symanzik:1983dc}, the chiral Lagrangian including all terms of $O(p^4,p^2m,m^2,p^2a,ma)$ is given in \cite{Rupak:2002sm}. The $O(a^2)$ contributions are constructed in \cite{Bar:2003mh,Aoki:2003yv}.
In the following we consider the case $N_f=2$ with degenerate quark mass $m$. 
The leading order Euclidean chiral Lagrangian in the continuum is given by \cite{Weinberg:1978kz,Gasser:1983yg}
\begin{equation}\label{eq1}
\mathcal{L}_2=\frac{F^2}{4}{\rm Tr}\left(\partial_\mu U\partial_\mu U^\dagger   \right)
-\frac{F^2Bm}{2}{\rm Tr}\left( U + U^\dagger\right).
\end{equation}
The pseudo Nambu-Goldstone modes are parametrized by the SU(2) field $U(x)=\exp\left(2i\xi(x)\right)/F$, 
and $F$, $B=\Sigma/F^2$ are the familiar LO couplings. The leading terms involving the lattice spacing are
\begin{eqnarray}
\mathcal{L}_a& = & \hat{a}W_{45}{\rm Tr}\left(\partial_\mu U\partial_\mu
    U^\dagger \right){\rm
    Tr}\left(U+U^\dagger\right)-\hat{a}\hat{m}W_{68}\left({\rm
      Tr}\left(U+U^\dagger\right)\right)^2,\\\label{eq2}
\mathcal{L}_{a^2}& = &\frac{F^2}{16}c_2 a^2 \left({\rm
      Tr}\left(U+U^\dagger\right)\right)^2,\label{eq3}
\end{eqnarray}
where $\hat{m}= 2Bm$ and $\hat{a}=2W_0 a$. $W_{45}$, $W_{68}$, $W_0$ and $c_2$ are new LECs. Note that the mass parameter $m$ in Eq. (\ref{eq2}) is the so-called \emph{shifted} mass \cite{Sharpe:1998xm}: besides the dominant additive mass renormalization proportional to $1/a$ it also contains the leading correction of $O(a)$. 

Currents and densities in WChPT can be constructed by a standard spurion analysis or by introducing source terms. Here we report the axial vector current and the pseudoscalar density including the leading $O(a)$ corrections \cite{Sharpe:2004ny,Aoki:2007es,Aoki:2009kt}:
\begin{equation}\label{eq3a}
A^a_{\mu,{\rm WChPT}} =  A^a_{\mu,{\rm cont}} \left\{ 1+\frac{4}{F^2}\hat{a}W_{45}{\rm Tr}(U+U^\dagger)\right\},\;\;\;\;\;\;\;\;\; P^a_{\rm WChPT}=   P^a_{\rm cont}   \left\{ 1+\frac{4}{F^2}\hat{a}W_{68}{\rm Tr}(U+U^\dagger) \right\}, 
\end{equation}
where 
\begin{equation}\label{eq3d}
A^a_{\mu,{\rm cont}}  =  i\frac{F^2}{2}{\rm Tr} \left(T^a(U^\dagger\partial_\mu U -U\partial_\mu U^\dagger)   \right),\;\;\;\;\;\;\;\;\;\;\;
P^a_{\rm cont}   =  i\frac{F^2B}{2} {\rm Tr}\left(T^a(U-U^\dagger)   \right),
\end{equation}
and $T^a$ are SU(2) generators normalized such that ${\rm Tr}(T^a T^b)=\delta^{ab}/2$.

\subsection{Power counting in infinite volume}
In WChPT there are two parameters which break explicitly the chiral symmetry, the quark mass $m$ (counted as $O(p^2)$) and the lattice spacing $a$. 
The power counting is determined by the relative size of these two parameters. In particular, one defines \cite{Sharpe:2004ny,Sharpe:2004ps} two different regimes: (i) the GSM \footnote{GSM stands for \emph{generically small masses}.} regime, where $m\sim a\Lambda^2_{\rm QCD}$ and (ii) the Aoki regime where $m\sim a^2\Lambda^3_{\rm QCD}$. In the Aoki regime lattice artifacts are more pronounced, and the $\mathcal{L}_{a^2}$ in Eq. (\ref{eq3}) enters already at LO. The pion mass at leading order is given by
\begin{eqnarray}
{\rm GSM\; regime} & : & M_0^2=2Bm,\\
{\rm Aoki\; regime} & : & M_0^2=2Bm-2c_2a^2. \label{maoki}
\end{eqnarray}
The sign of $c_2$ governs the phase diagram of the theory \cite{Sharpe:1998xm}.

\subsection{Power counting in the $\epsilon$-regime}
In the $\epsilon$-regime the chiral limit is approached by keeping  $\mu=m\Sigma V\lesssim O(1)$; this corresponds to the situation where the pion wavelength is much larger than the linear size of the box.
The main effect of formulating the effective theory in this regime is that the pion zero-mode becomes non-perturbative and its contribution has to be treated exactly.  This is achieved by factorizing the pseudo Nambu-Goldstone boson fields as
\begin{equation}
U(x)=\exp\left(\frac{2i}{F}\xi(x)   \right)U_0,
\end{equation}
where the constant $U_0\in$ SU(2) represents the collective zero-mode. The non-zero modes $\xi$ can still be treated perturbatively. 
The $\epsilon$-regime requires a reorganization of the perturbative series: in the continuum, this corresponds to taking the quark mass of order $m\sim O(\epsilon^4)$. 

In order to extend the WChPT to the $ \epsilon$-regime, one has to assign a relative power counting of the lattice spacing $a$ with respect to the quark mass $m$. By assuming that the quark mass can be considered of order $m\sim O(\epsilon^4)$ also in  WChPT 
\footnote{While this is a natural choice in the GSM regime, the situation in the Aoki regime can be more subtle. See \protect\cite{Bar:2008th} for a more detailed discussion on this issue. }, we obtain 
\begin{eqnarray}
{\rm{GSM\; regime}} &: & m\sim O(a\Lambda^2_{\rm QCD})\;\rightarrow\; a\sim O(\epsilon^4),     \\
{\rm{Aoki\; regime}} &: & m\sim O(a^2\Lambda^3_{\rm QCD})\;\rightarrow\; a\sim O(\epsilon^2).   
\end{eqnarray}
In addition, the $\epsilon$- expansion allows us to introduce an intermediate counting between the GSM and the Aoki regime: we can define the 
GSM$^*$ regime, where $a\sim O(\epsilon^3)$.\\
We compute mesonic two-point functions within the WChPT in the $\epsilon$-regime. In particular, we give explicit results for the pseudoscalar and axial time correlators, 
\begin{equation}
\delta^{ab}C_{PP}(t)=\int d^3\vec{x}\langle P^a(x) P^b(0)\rangle,\;\;\;\; \delta^{ab}C_{AA}(t)=\int d^3\vec{x}\langle A_0^a(x) A_0^b(0)\rangle.
\end{equation}
Currents and densities are defined in Eq. (\ref{eq3a}); the subscript ``WChPT'' is now omitted.
With our power counting, we find that in the GSM regime lattice corrections enter only at NNLO, while in the Aoki regime effects of lattice artifacts show up already at LO.

In the following we concentrate on the intermediate GSM$^*$ regime, where corrections appear at NLO. At this order, they are given only by the term $\mathcal{L}_{a^2}$ in Eq. (\ref{eq3}).
 This suppression comes from the fact that the lattice spacing corrections in the chiral effective theory action and in the effective operators are either quadratic in $a$ or they come with an additional power of either $m$ or $p^2$. 
Notice that this is valid also for unimproved Wilson fermions: a non-perturbative $O(a)$ improvement removes corrections due to $\mathcal{L}_{a}$ as well as the $O(a)$ terms in the operators, i.e. acts only on terms which are subleading in our power counting. 

%%%%%%%%%%%%%%%%%%%%%%%%%%%%%%%%%%%%%%%%%%%%%%%%%%%%%%%%%%%%%%%%%%%%%%%%%%%%%%%%%%%%%%%%%%%%%%%%%%
\subsection{Leading corrections in the GSM$^*$ regime}
The continuum pseudoscalar and axial correlators at NLO in the $\epsilon$-expansion can be written as \cite{Hansen:1990un}
\begin{equation}\label{eq5}
C_{PP,AA\;\rm{ct}}(t)=a_{P,A} + b_{P,A} h_1(t/T),
\end{equation}
with a parabolic time dependence given by the function $h_1(\tau)=\frac{1}{2}\left[\left(| \tau|-\frac{1}{2}  \right)^2 -\frac{1}{12} \right]$.\\
For $N_f=2$ the coefficients $a_{P,A},b_{P,A}$ explicitly read \cite{Hansen:1990un}
\begin{eqnarray}
a_P & = & \frac{L^3}{2}\frac{\Sigma^2_{\rm eff}}{\mu_{\rm eff}} \frac{I_2(2\mu_{\rm eff})}{I_1(2\mu_{\rm eff})},\;\;\;\;\;\;\;\;\;\;\;\;\;\;\;\;\;\; b_P =  \frac{T\Sigma^2}{2F^2} \left[2-\frac{1}{\mu}\frac{I_2(2\mu)}{I_1(2\mu)}   \right],                      \label{eq7}\\
a_A & = & -\frac{F^2}{T} \left[1-\frac{I_2(2\mu_{\rm eff})}{\mu_{\rm eff}I_1(2\mu_{\rm eff}) }    \right]  -\frac{2\beta_1}{T\sqrt{V}}  \left[2-\frac{1}{\mu}\frac{I_2(2\mu)}{I_1(2\mu)}   \right]      +\frac{2T}{V}k_{00}\frac{I_2(2\mu)}{\mu I_1(2\mu)},    \nonumber  \\
b_A & = &  -\frac{2T}{V}  \frac{\mu I_2(2\mu)}{I_1(2\mu)}.  \label{eq7a}
\end{eqnarray}
$I_1$, $I_2$ are modified Bessel functions of the first kind; $\beta_1$ and $k_{00}$ are so-called shape factors \cite{Hasenfratz:1989pk,Hansen:1990un}, which depend only on the geometry of the finite box.
$\Sigma_{\rm eff}$ is the quark condensate at one loop \cite{Gasser:1986vb}
\begin{equation}
\Sigma_{\rm eff}=\Sigma\left( 1+\frac{3}{2F^2}\frac{\beta_1}{\sqrt{V}}  \right),
\end{equation}
and $\mu_{\rm eff}=m\Sigma_{\rm eff} V$. As anticipated, the continuum NLO expressions contain only the LO LECs $\Sigma$ and $F$. The first non-trivial modification of the continuum NLO results appear in the GSM$^*$ regime and it is due to $\mathcal{L}_{a^2}$ only. 
In this case we can write down the full NLO correlators in WChPT as
\begin{equation}
C_{PP,AA}(t)= C_{PP,AA\;\rm{ct}}(t)+ C_{PP,AA\; a^2}(t).
\end{equation}
By performing the explicit computation (see \cite{Bar:2008th} for the full details) it turns out that the corrections $C_{PP,AA\; a^2}(t)$ are time-independent, hence affect only the constant part of the correlators. In particular we obtain:
\begin{equation}
C_{PP a^2}= \frac{L^3\Sigma^2}{2}\rho\Delta_{a^2},\;\;\;\;C_{AA a^2}= \frac{F^2}{T}\rho\Delta_{a^2},
\end{equation}
where
\begin{equation}\label{eq9}
\Delta_{a^2}=\frac{5\mu I_1^2(2\mu)-10I_1(2\mu)I_2(2\mu)-3\mu I_2^2(2\mu)}{2\mu^3I_1^2(2\mu)},
\end{equation}
and $\rho=F^2c_2a^2V$ is the dimensionless LEC which parametrizes the $O(a^2)$ correction.

It is useful to compute the leading $O(a^2)$ corrections to the PCAC quark mass:
\begin{equation}
m_{\rm PCAC}=m\left[1+\rho \left(\frac{2}{\mu^2} -\frac{I_1(2\mu)}{\mu I_2(2\mu)}   \right)    \right].
\end{equation}
It is now possible to express the correlators $C_{PP,AA}(t)$ as a function of $\tilde{\mu}=m_{PCAC}\Sigma V$; the result is
\begin{equation}
C_{PP}(t)= C_{PP\rm{ct}}(t)+ \frac{L^3\Sigma^2}{2}\rho\tilde\Delta_{a^2}, \;\;\;\;\;C_{AA}(t)= C_{AA\rm{ct}}(t)+\frac{F^2}{T}\rho\tilde\Delta_{a^2},
\end{equation}
where the continuum correlators are as in Eq. (\ref{eq5}), but with the replacements $\mu\rightarrow \tilde\mu$, $\mu_{\rm eff}\rightarrow \tilde\mu_{\rm eff}=m_{\rm PCAC}\Sigma_{\rm eff} V$ and
\begin{equation}
\tilde\Delta_{a^2}=\frac{4\tilde\mu^2I_1^3(2\tilde\mu)-11\tilde\mu I_1^2(2\tilde\mu)I_2(2\tilde\mu)+2(3-2\tilde\mu^2)I_1(2\tilde\mu)I_2^2(2\tilde\mu)+5\tilde\mu I_2^3(2\tilde\mu)     }{2\tilde\mu^3I_1^2(2\tilde\mu)I_2(2\tilde\mu)}.
\end{equation}
Other correlation functions can be computed along the same line. 
%%%%%%%%%%%%%%%%%%%%%%%%%%%%%%%%%%%%%%%%%%%%%%%%%%%%%%%%%%%%%%%%%%%%%%%%%%%%%%%%%%%%%%%%%%%%%%%%%%%%%%%%%
\begin{figure}
\begin{minipage}{7.7cm}
\includegraphics[width=8.3cm]{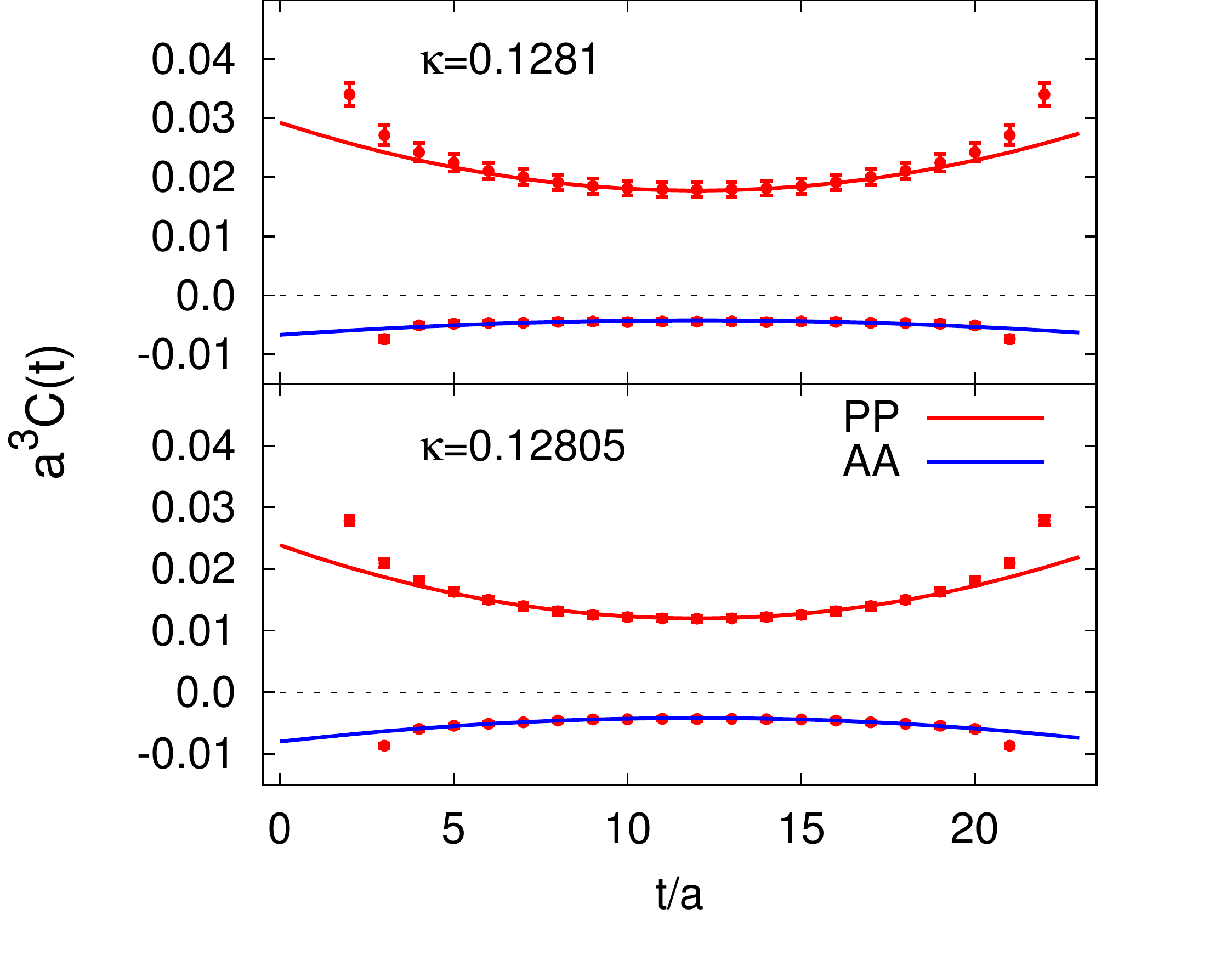}
\end{minipage}
\begin{minipage}{7.7cm}
\includegraphics[width=8.3cm]{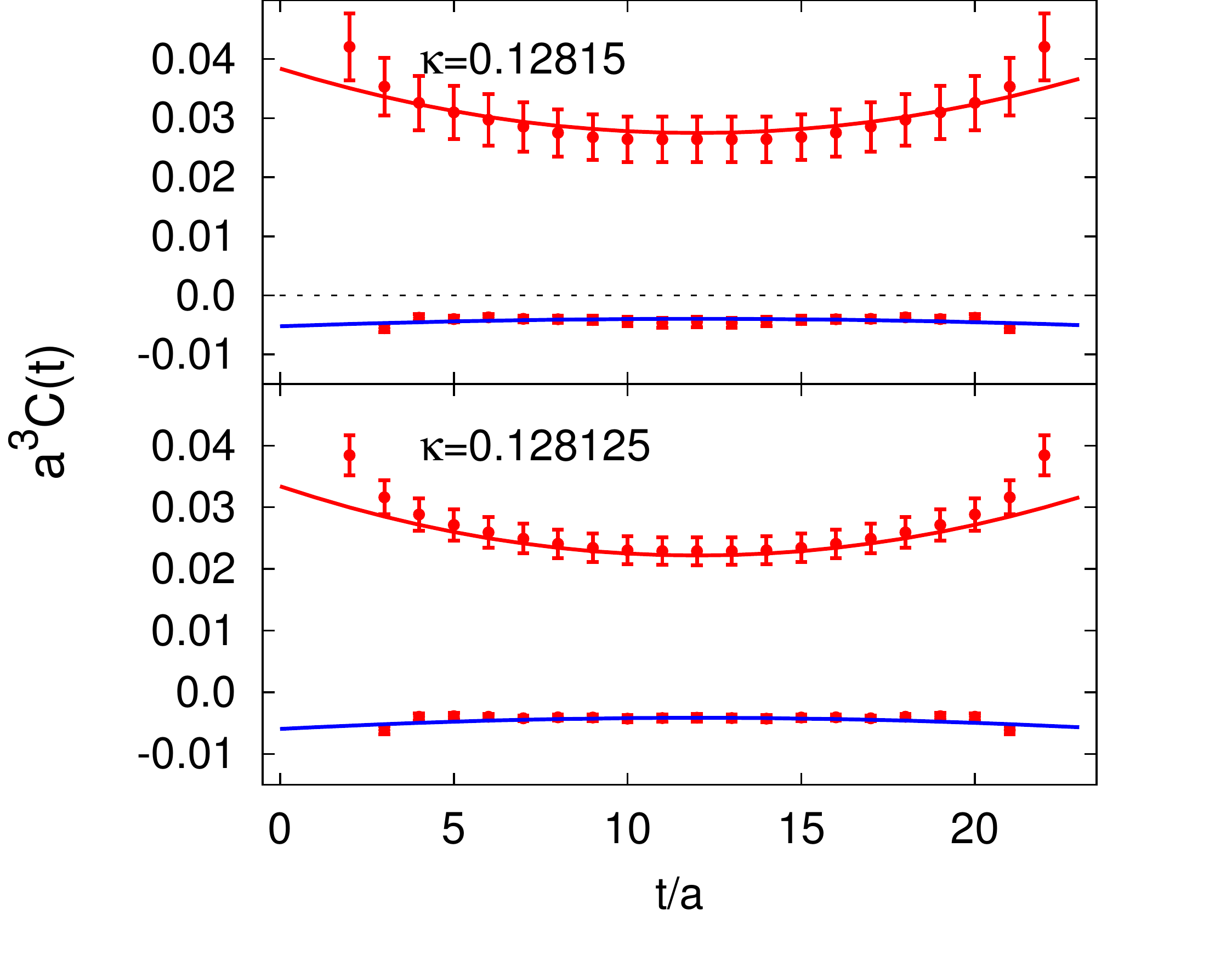}
\end{minipage}
\caption{Fit of the WChPT predictions to lattice data. 
All data points within the fit range of $t/a\in[6,18]$ for the four sea quark masses are included in the combined fit. 
The hopping parameter $\kappa=(0.128150,0.128125,0.1281,0.128050)$ corresponds to $am_{\rm PCAC}=(0.0019(4),0.0024(3),0.0030(3),0.0044(3))$ respectively.
The axial vector correlator is multiplied by a factor 50 for better visibility.  }\label{fig1}
\end{figure}
\section{Reanalysis of lattice data and conclusions}
These predictions from WChPT at NLO can be tested against lattice data generated in \cite{Hasenfratz:2008ce}, where pseudoscalar and axial correlators have been computed on an ensemble with $N_f=2$ flavors of dynamical improved NHYP Wilson fermions \cite{Hasenfratz:2007rf}. 
At a lattice spacing $a\simeq 0.115$ fm, two lattice extents are considered, 
$L_1=16a\simeq 1.84$ fm and $L_2=24a\simeq 2.8$ fm. Quark masses approach the $\epsilon$-regime, with $\tilde\mu\simeq 0.7-2.9$ for the volume $V_1=L_1^4$ and  $\tilde\mu\simeq 2.1-5.0$ for the volume $V_2=L_2^4$.
In the GSM$^*$ regime, we have only the additional LEC $c_2$ with respect to the continuum case. Notice that its value will depend on the particular discretized action which is used. 
We simultaneously fit the two correlators for all      available quark masses; for the volume $V_2$, a fit in the range $t\in[6,18]$ gives
\begin{equation}\label{res}
\left[\Sigma^{\overline{\rm MS}}(\mu=2\;{\rm GeV})\right]^{1/3} =249(4)\; {\rm MeV},\;\;\; F=88(3) \; {\rm MeV},\;\;\;c_2=0.02(8)\;{\rm GeV}^4.
\end{equation}
The data, along with the theoretical curves, are shown in Fig. \ref{fig1}. The errors from the renormalization factors $Z_A$, $Z^{\overline{\rm MS}}_P(\mu=2\;{\rm GeV})$ computed in \cite{Hasenfratz:2008ce} are not included in the uncertainties of the LECs. Varying the time range of the fit does not give significant differences for the LECs, as long as $t_{\rm min}/a >4$. Also discarding the heaviest mass does not change the results of Eq. (\ref{res}) within the statistical errors. 
The smallest volume $V_1$ yields values which are consistent with Eq. (\ref{res}), but the large $\chi^2$ of the fit may indicate that NLO formulae are no longer applicable. 
The values of $F$ and $\Sigma$ are compatible with other determinations \cite{Necco:2009cq}, while the value of $c_2$ is compatible with zero. 
 A continuum fit (with $c_2=0$) yields virtually unchanged values for $F$ and $\Sigma$, showing that cut-off effects do not impact the extraction of the LECs beyond the level of the statistical uncertainties.\\
This is a very encouraging result: simulations with Wilson fermions in the $\epsilon$-regime are feasible 
and seem to be a viable alternative to dynamical simulations with GW fermions. 
Lattice computations on a wide range of lattice spacings and volumes would of course be very useful to test if the predicted NLO scaling is verified. 
The results derived here can be generalized in various ways, for example to the case with a twisted mass term or to an arbitrary number of flavors.

%\bibliography{lattice}        %or whatever your .bib file is
%   \bibliographystyle{h-elsevier}   %if you use h-elsevier.bst

\end{document}